\documentclass[floats,prb,twocolumn,amsmath,showpacs]{revtex4}

\usepackage{epsfig}
\usepackage{colordvi}
\usepackage{graphicx}

\def\k{{\bf k}}
\def\E{{\cal E}}

\def\R{{\bf R}}
\def\P{{\bf P}}
\def\g{{\bf g}}

\begin{document}

\marginparwidth 3.1in
\marginparsep 0.5in
\def\dvm#1{\marginpar{\small DV: #1}}
\def\xwm#1{\marginpar{\small XW: #1}}

\title{First-principles perturbative computation of dielectric and
Born charge tensors in finite electric fields}

\author{Xinjie Wang and David Vanderbilt}
\affiliation{Department of Physics and Astronomy, Rutgers University,
	Piscataway, NJ 08854-8019}
\date{\today}

\begin{abstract}
We present a perturbative treatment of the response properties of
insulating crystals under a dc bias field, and use this to study
the effects of such bias fields on the Born effective charge tensor
and dielectric tensor of insulators. We start out by expanding a
variational field-dependent total-energy functional with respect
to the electric field within the framework of density-functional
perturbation theory. The second-order term in the expansion of the
total energy is then minimized with respect to the first-order
wave functions, from which the Born effective charge tensor
and dielectric tensor are easily computed.  We demonstrate an
implementation of the method and perform illustrative calculations
for the III-V semiconductors AlAs and GaAs under finite bias field.
\end{abstract}


\maketitle

\section{Introduction}

The dielectric tensor and Born (or dynamical) effective charge tensor are of
fundamental importance in understanding and modeling the response
of an insulator to an electric field.\cite{lst}  They give,
respectively, the first-order polarization and atomic force
appearing in response to a first-order change in the
macroscopic electric field.  While one is most often interested
in evaluating these response tensors at zero field, there
is increasing interest in finite-field properties.  For example,
the study of bulk ferroelectrics\cite{sai02,fu03,dieguez06} and of
ferroelectric films\cite{antons05} and superlattices\cite{dieguez06,wu06}
in finite field, and of lattice vibrations in polar crystals in
finite field,\cite{wang06} have recently generated interest.
While it may sometimes be reasonable to model the dielectric behavior
by assuming that the dielectric and Born effective charge tensors
have a negligible dependence on the bias field, it is important
to be able to quantify such approximations and to compute the
field dependence when it is physically important to do so (e.g.,
for describing non-linear optical phenomena such as
second-harmonic generation).

Density-functional perturbation theory (DFPT)\cite{dfpt1,dfpt2}
provides a powerful tool for calculating the second-order
derivatives of the total energy of a periodic solid with respect
to external perturbations such as atomic sublattice displacements
or a homogeneous electric field.  In contrast to the case of
sublattice displacements, for which the perturbing potential
remains periodic, the treatment of homogeneous electric fields
is subtle because the corresponding potential acquires a term
that is linear in real space, thereby breaking the translational
symmetry and violating the conditions of Bloch's theorem.
For this reason, electric-field perturbations have often been
studied in the past using the long-wave method, in which the
linear potential resulting from the applied electric field is
obtained by considering a sinusoidal potential in the limit that
its wave vector goes to zero.  In this approach, however,
the response tensor can only evaluated at zero electric field, and
it also requires as an ingredient the calculation of the
derivatives of the ground-state wave functions with respect to
wave vector.

Recently, Nunes and Gonze introduced an electric-field-dependent
energy functional expressed in terms of the Berry-phase
polarization.\cite{nunes01}  This approach was initially introduced
in order to provide an alternative framework for the DFPT treatment
of electric-field perturbations (evaluated at zero field) in
which the long-wave method is entirely avoided.  More recently,
it has been pointed out that the Nunes-Gonze functional could
also serve as the basis for a calculation of the {``ground-state''}
properties in {\it finite} electric field.\cite{souza02,umari02}
(Here, the phrase ``ground state'' is used advisedly; because of
Zener tunneling, the state of interest is actually a long-lived
resonance.\cite{souza04})  In this approach, the energy functional
is minimized with respect to a set of field-polarized Bloch
functions that form a natural representation of the one-particle
density matrix even though they are no longer eigenstates of
the Hamiltonian.\cite{souza02,souza04}  The introduction of this
approach has also made possible the calculation of the dielectric
and Born effective charge tensors at finite electric
fields using finite-difference methods.\cite{souza02}

In a recent paper\cite{wang06} we developed a perturbative method,
within this framework, for computing the phonon properties of insulators
at finite electric fields.  The starting point was the Nunes-Gonze
electric-field-dependent energy functional, which represents the effect
of the electric field by including its coupling to the Berry-phase
polarization.\cite{nunes01} This total-energy functional was expanded up
to second order in atomic displacements. The linear response
of the field-polarized Bloch functions to the atomic displacements was
obtained by minimizing the second-order term in the expansion of the
total-energy functional with respect to the first-order changes in the
Bloch functions.  Finally, the force-constant matrix was
constructed based on these first-order Bloch functions.
This method provides a tractable and efficient computational
scheme for computing phonon properties at finite electric field,
and suggests that a similar treatment of other response properties
of insulators in finite electric field should be possible.

In this paper, we follow the approach of
Ref.~\onlinecite{wang06} to develop a method for computing the dielectric
and Born effective charge tensors at finite electric field.
Again using the Nunes-Gonze energy functional,\cite{nunes01}
we compute the first-order responses of the electronic wave functions
to a small change in the electric field.  We then use these to construct
the second-order derivatives of the total energy with respect to
electric field (to give the dielectric tensors evaluated at non-zero field)
and the mixed derivatives with respect to electric field and atomic
sublattice displacement (to give the Born effective charge tensors
evaluated at non-zero field).

The paper is organized as follows. In Sec.~\ref{sec:method}, the second-order
perturbation expansion of the total energy functional with respect to
electric fields is derived and the steepest-descent directions are
identified.  The expressions for computing the dielectric and
Born effective charge tensors are also given.
In Sec.~\ref{sec:test}, we describe the implementation of the approach in
the {\tt ABINIT} code package,\cite{abinit} and present test calculations
for the III-V semiconductors AlAs and GaAs.  (Since we are mainly
interested in the purely electronic effects here, we do not include the
strains or sublattice displacements that might occur in response to the
electric field; these could easily be included by employing
structural-relaxation methods at finite field.\cite{souza02})
By comparing with the
results of finite-differences calculations, we demonstrate the correctness
of the new formulation and the internal consistency of the theory.
A brief summary and conclusion are presented in Sec.~\ref{sec:sum}.

\section{method}
\label{sec:method}

\subsection{Perturbation expansion of the enthalpy functional}
\label{sec:pertexp}

We start from the electric enthalpy functional\cite{nunes01,souza02}
\begin{equation}
F[\R;\psi;\E]=E_{\rm KS}[\R;\psi]
-\Omega \E\cdot \P[\psi] \;,
\label{eq:Ftotal}
\end{equation}
where $\R$, $\E$, $\Omega$ and $\P$ are, respectively, the
atomic coordinates, the electric field, the cell volume, and
the macroscopic polarization, $E_{\rm KS}$ is Kohn-Sham energy
functional at zero electric field, and atomic units are used
throughout.  After minimizing this functional,
the field-polarized Bloch functions $\psi$ may be regarded as
depending implicitly on the electric field $\E$.
Our treatment of this functional will parallel the treatment
given in our previous Ref.~\onlinecite{wang06}.

In the present
case, we take the electric field $\E$ to consist of two parts,
a finite part $\E^{(0)}$ and a small variation $\delta\E$. In
the following, we consider the perturbation expansion of the
functional in Eq.~(\ref{eq:Ftotal}) with respect to the small
variation $\delta\E$ under the orthonormality constraints
\begin{eqnarray}
	\langle \psi_{m\k}|\psi_{n\k}\rangle=\delta_{mn} \;.
\end{eqnarray}
The wave functions are to be relaxed, subject to these constraints,
in such a way as to minimize the electric enthalpy functional
\begin{eqnarray}
	F=F_{\rm KS}+F_{\rm BP}+F_{\rm LM}\;,
	\label{eq:ftotal}
\end{eqnarray}
where $F_{\rm KS}=E_{\rm KS}$ is the Kohn-Sham energy (as it would
be calculated at $\E=0$), $F_{\rm BP} =-\Omega {\E} \cdot \P$
contains the coupling of the Berry-phase polarization $\P$
to the electric field, and the constraint is implemented by
the inclusion of the Lagrange-multiplier term $F_{\rm LM}$.
The first and last of these terms are given by
\begin{equation}
F_{\rm KS}=\frac{f}{N_{k}}\sum_{\k n}^{\rm occ}
   \langle \psi_{n\k}|T+v_{{\rm ext}}|\psi_{n\k}\rangle +E_{\rm Hxc}[n]
\label{eq:Fbiga}
\end{equation}
and
\begin{equation}
F_{\rm LM} = -\frac{f}{N_k}\sum_{\k,mn}^{\rm occ} \Lambda_{\k,mn}
(\langle \psi_{m\k}|\psi_{n\k}\rangle-\delta_{mn})
\label{eq:Fbigc}
\end{equation}
where $f$ is the spin degeneracy (normally $f$=2), $N_{k}$ is the number of
$k$-points, and $\Lambda_{\k,mn}$ is the matrix of Lagrange multipliers.
In a notation similar to that of Ref.~\onlinecite{wang06}, the second
term may be written as
\begin{equation}
F_{\rm BP}=-\frac{ef}{2\pi}\sum_{i=1}^{3}\frac{{\cal E} \cdot {\bf a}_i}
{N_{\perp }^{(i)}} \sum_{\k} D_{\k,\k+\g_i} \;.
\label{eq:Fbigb}
\end{equation}
Here ${\bf a}_i$ are the three primitive real-space lattice vectors,
and the mesh of $N_k$ $k$-points is defined by mesh vectors
${\bf g}_i={\bf b}_i/N^{(i)}$ where ${\bf b}_i$ is the reciprocal
lattice vector dual to ${\bf a}_i$.  Thus, $N_k=N^{(1)}N^{(2)}N^{(3)}$,
and we also define $N_\perp^{(i)}=N_k/N^{(i)}$ as the
number of $k$-point strings running in direction $i$.  Finally,
\begin{equation}
D_{\k\k'}=\textrm{Im}\ln\det S_{\k\k'}
\end{equation}
where the overlap matrix is defined as
\begin{eqnarray}
	(S_{\k\k'})_{mn}&=& \langle u_{m\k}|u_{n\k'}\rangle\;.
\label{eq:overlap}
\end{eqnarray}

In order to obtain the desired response properties, we now wish to expand
the finite-field enthalpy functional $F_{\rm KS}$ up to second
order in the electric field. We shall assume for the moment that
the electric field is applied in Cartesian direction $\alpha$ only.
The expansion of $F_{\rm KS}$ with respect to atomic displacements
was already obtained in Ref.~\onlinecite{wang06}, and the expansion with
respect to electric field can be carried through in a very similar way.
Indeed, the second-order expansions of $F_{\rm KS}$ and $F_{\rm LM}$
can essentially be transcribed from Ref.~\onlinecite{wang06} with
the first-order wave functions with respect to displacement replaced
here by the first-order wave functions with respect to electric field,
giving
\begin{eqnarray}
\nonumber
F^{(2)}_{\rm KS}&=& \frac{1}{2}\,
	\frac{\partial^2 F_{\rm KS}}{\partial \E_{\alpha}^2}\\
	\nonumber
		&=&\frac{f}{N_k}\sum_{\k}^{}\sum_{n=1}^{occ}
\langle u_{n{\k}}^{\E_{\alpha}}|
T+v^{}_{ext}|u_{n{\k}}^{\E_{\alpha}}\rangle \\
&&\qquad\qquad\qquad+E_{\rm Hxc}^{\E_{\alpha}\E_{\alpha}}
\label{eq:fks}
\end{eqnarray}
and
\begin{eqnarray}
F^{(2)}_{\rm LM}&=&-\frac{f}{N_k}\sum_{\k,n}^{\rm occ}
\Lambda_{\k,nn}^{(0)}
\langle u_{n{\k}}^{\E_{\alpha}}|
u_{n{\k}}^{\E_{\alpha}}\rangle
\;.
\label{eq:flm}
\end{eqnarray}
As in Ref.~\onlinecite{wang06}, terms that can be eliminated by use
of the ``$2n+1$ theorem'' (e.g., $\langle u_{n{\k}}^{\E_{\alpha}\E_{\alpha}}
|T+v^{}_{ext}|u_{n{\k}}^{(0)}\rangle$) have been dropped.  The
the first-order wave functions are
\begin{equation}
|u_{n{\k_j}}^{\E_{\alpha}}\rangle=\frac{\partial
|u_{n{\k_j}}\rangle}{\partial \E_{\alpha}}
\label{eq:1stwf}
\end{equation}
and the second-order $E_{\rm Hxc}$ are
\begin{equation}
E_{\rm Hxc}^{\E_{\alpha}\E_{\alpha}}=\frac{\partial^2 E_{\rm Hxc}}
{2\partial \E_{\alpha}\partial\E_{\alpha}}\;.
\end{equation}
In these and subsequent equations, the partial derivatives indicate
that the {\it structural} coordinates $\R$ are being held fixed (while,
however, the {\it wave functions} $|u_{n\k}\rangle$ are allowed to vary).

The second-order expansion of $F_{\rm BP}$ with respect to electric field
requires somewhat more care.  We find
\begin{eqnarray}
 \nonumber
	F_{\rm BP}^{(2)} &=& \frac{1}{2}\,
            \frac{\partial^2 F_{\rm BP}}{\partial \E_{\alpha}^2}
	\\ \nonumber
         &=&-\frac{\Omega}{2}\, \frac{\partial^2 (\E \cdot \P)}
	{\partial \E_{\alpha}^2}\\
	&=&-\Omega(\hat{\bf e}_{\alpha} \cdot \P^{\E_\alpha}_{}+\E^{(0)}
	\cdot \P^{\E_\alpha \E_\alpha}_{})\;,
	\label{eq:fbp}
\end{eqnarray}
where $\hat{\bf e}_{\alpha}$ is the unit vector along Cartesian direction
$\alpha$.  The first term in the last line of Eq.~(\ref{eq:fbp}) is
special to the case of the electric-field perturbation, while the second term
can be derived in close correspondence to the case of displacement
perturbations in Ref.~\onlinecite{wang06}.
The first-order variation of $\P$ with field $\E_\alpha$ is
\begin{eqnarray}
  \P^{\E_\alpha}_{} = -\frac{ef}{2\pi\Omega}\sum_{i=1}^{3}
         \frac{{\bf a}_{i}}{N_{\perp }^{(i)}} \sum_\k D^{(1)}_{\k,\k+\g_i}
  \label{eq:ep1a}
\end{eqnarray}
and its second-order variation is
\begin{eqnarray}
  \P^{\E_\alpha\E_\alpha}_{} = -\frac{ef}{4\pi\Omega}\sum_{i=1}^{3}
         \frac{{\bf a}_{i}}{N_{\perp }^{(i)}} \sum_\k D^{(2)}_{\k,\k+\g_i} \;,
  \label{eq:ep2a}
\end{eqnarray}
where
\begin{eqnarray}
      D^{(1)}_{\k,\k+\g_i} =
  \textrm{Im}{\rm Tr}\left [S_{\k,\k+\g_i}^{(1)}Q_{\k+\g_i,\k}\right ]
  \label{eq:ep1b}
  \end{eqnarray}
and
\begin{eqnarray}
      D^{(2)}_{\k,\k+\g_i} &=&
  \textrm{Im}{\rm Tr}\Big [2{ S}_{\k,\k+\g_i}^{(2)}Q_{\k+\g_i,\k}
\nonumber\\
&&-
{S}_{\k,\k+\g_i}^{(1)}Q_{\k+\g_i,\k}{S}_{\k,\k+\g_i}^{(1)}Q_{\k+\g_i,\k}
\Big]
  \;.
  \label{eq:ep2b}
  \end{eqnarray}
In these equations,
`Tr' indicates a trace of the bracketed matrix over band indices,
and $Q$, $S^{(1)}$, and ${S}^{(2)}$ are defined with respect
to the series expansion of the overlap matrix via
\begin{eqnarray}
	S_{\k\k'}(\E_{\alpha})=S_{\k\k'}^{(0)}+\E_{\alpha} S_{\k\k'}^{(1)}+
 \E_{\alpha}^2 S_{\k\k'}^{(2)}+...
\end{eqnarray}
and
\begin{eqnarray}
	Q_{\k\k'}&=& [S_{\k'\k}^{(0)}]^{-1} \;.
\end{eqnarray}
The first- and second-order expansions of the overlap matrix take the form
\begin{equation}
{ S}_{\k,\k',{mn}}^{(1)}=
\langle u_{m\k}^{\E_{\alpha}}|u_{n\k'}^{(0)}\rangle+
\langle u_{m\k}^{(0)}|u_{n\k'}^{\E_{\alpha}}\rangle
\end{equation}
and
\begin{equation}
{S}_{\k,\k',{mn}}^{(2)}=
\langle u_{m\k}^{\E_{\alpha}}|u_{n\k'}^{\E_{\alpha}}\rangle \;.
\end{equation}
In the last equation above, terms like
$\langle u_{m\k}^{\E_{\alpha}\E_{\alpha}}|u_{n\k'}^{(0)}\rangle$
have again been dropped by virtue of the ``$2n+1$ theorem.''

\subsection{First-order wave functions with respect to electric-field
perturbation}
\label{sec:wfpert}

The second-order term in the expansion of the energy functional,
given by the sum
$F^{(2)}=F^{(2)}_{\rm KS}+F^{(2)}_{\rm BP}+F^{(2)}_{\rm LM}$
of the expressions in Eqs.~(\ref{eq:fks}), (\ref{eq:fbp}), and
(\ref{eq:flm}) respectively, is minimized with respect to
the first-order wave functions $|u_{n\k}^{\E_{\alpha}}\rangle$ using
standard conjugate-gradient methods.  The steepest-descent direction
is obtained from the gradient of $F^{(2)}$ with respect to
$\langle u_{n\k}^{\E_{\alpha}}|$, whose contributions take the form
\begin{eqnarray}
	\frac{\delta  F^{(2)}_{\rm KS}}{\delta u^{\E_{\alpha}*}_{n\k}}&=&
	\frac{f}{N_k}\left [ \left (T+v^{(0)}_{ext}\right )
	|u_{n\k}^{\E_{\alpha}}\rangle
  +\frac {\delta E_{\rm Hxc}^{\E_{\alpha}\E_{\alpha}}}
  {\delta u^{\E_{\alpha}*}_{n\k}}\right ] \;,
\end{eqnarray}
\begin{eqnarray}
\nonumber
 \frac{\delta  F^{(2)}_{\rm BP}}{\delta u^{\E_{\alpha}*}_{n\k}}&=&
  \frac{ief}{4\pi}\sum_{i=1}^{3}
 \frac{{\E^{(0)}} \cdot {\bf a}_i}{N_{\perp }^{(i)}}
\big(\,|{\cal C}_{m\k,\k+\g_i}\rangle-|{\cal C}_{m\k,\k-\g_i}\rangle\,\big)\\
\nonumber
&+& \frac{ief}{4\pi}\sum_{i=1}^{3}
\frac{{\hat {\bf e}}_{\alpha} \cdot {\bf a}_i}{N_{\perp }^{(i)}}
\big(\,|{\cal D}_{m\k,\k+\g_i}\rangle-|{\cal D}_{m\k,\k-\g_i}\rangle\,\big)
\;, \\
\end{eqnarray}
and
\begin{equation}
	\frac{\delta  F^{(2)}_{\rm LM}}{\delta u^{\E_{\alpha}*}_{n\k}}
	=\frac{f}{N_k}\epsilon_{n\k}^{(0)}|u_{n{\k}}^{\E_{\alpha}}\rangle \;.
\end{equation}
Here
\begin{equation}
{\cal C}_{m\k\k'}= \left(\,
|u_{\k'}^{\E_{\alpha}}\rangle Q^{\phantom{()}}_{\k'\k}
- |u_{\k'}^{(0)}\rangle Q^{\phantom{()}}_{\k'\k}S_{\k\k'}^{(1)}
                        Q^{\phantom{()}}_{\k'\k} \,\right)_{m}
\;,
\end{equation}
\begin{equation}
{\cal D}_{m\k\k'}= \left(\,
|u_{\k'}^{(0)}\rangle Q^{\phantom{()}}_{\k'\k}
\,\right)_{m} \;,
\end{equation}
and $\epsilon_{n\k}^{(0)}$ is the diagonal zero-order matrix of
Lagrange multipliers.
Convergence of the conjugate-gradient procedure yields
a set of first-order wave functions $|u_{n\k}^{\E_{\alpha}}\rangle$.
These then become the essential ingredients for constructing the
dielectric and Born charge tensors as discussed below.

\subsection{Dielectric permittivity tensor}
\label{sec:dielperm}

The dielectric permittivity tensor can be written as
\begin{eqnarray}
	\epsilon_{\alpha \beta}^{\infty}=\delta_{\alpha \beta}+{4\pi}
	\chi_{{\alpha}{\beta}}
	\label{eq:dieinf}
\end{eqnarray}
where the electric susceptibility tensor $\chi_{{\alpha}{\beta}}$
at a {\it finite} electric field is defined as
\begin{eqnarray}
  \nonumber
	\chi_{\alpha \beta}&=&-\frac{1}{\Omega}\frac{\partial^2
	F_{}(\E)}
	{ {\partial \E_{\alpha}\partial \E_{\beta}}}\bigg |_{\E=\E^{(0)}}\\
	&=&\frac{\partial
	P_{\alpha}}
	{ {\partial \E_{\beta}}}\bigg |_{\E=\E^{(0)}}=
	{\hat {\bf e}_{\alpha}}\cdot \P_{}^{\E_\beta}\;.
	\label{eq:dies}
\end{eqnarray}
The derivative $\P_{}^{\E_\beta}$ of the polarization with respect
to electric field is already given by Eq.~(\ref{eq:ep1a}).
Since the first-order wave functions
$| u_{n\k}^{\E_{\alpha}}\rangle$ have already been
obtained in Sec.~\ref{sec:wfpert}, it is straightforward to
evaluate Eq.~(\ref{eq:dies}) and thus obtain the polarizability
and permittivity.

The dielectric responses above are the static responses computed with
atomic coordinates frozen.  That is, they correspond to the dielectric
response that would be measured at frequencies low compared to electronic
frequencies but high compared to any infrared-active phonon modes.
The true static susceptibility could be computed by including the
lattice displacements (and, if appropriate, the piezoelectric
strains) using, e.g., the methods of Ref.~\onlinecite{wu05}.

\subsection{Born effective charge tensor}
\label{sec:zstar}

The electronic contribution to the Born effective
charge tensor at finite electric field takes the form
\begin{eqnarray}
	Z^*_{\kappa,\alpha \beta}&=&-\frac{\partial^2 F(\E)}{\partial \E_{\alpha}
	\partial \tau_{\kappa,\beta}}\bigg |_{\E=\E^{(0)}}.
	\label{eq:bec}
\end{eqnarray}
This expression can be calculated equivalently in two different ways.
First, introducing the force
$f_{\kappa,\alpha}=-\partial F(\E)/\partial\tau_{\kappa,\alpha}$
acting on atom $\kappa$ in direction $\alpha$, it can be written as
\begin{eqnarray}
Z^*_{\kappa,\alpha \beta}=	\frac{\partial
f_{\kappa,\beta}}{\partial \E_{\alpha}}\;.
\end{eqnarray}
Using the Hellmann-Feynman theorem, the expression for the force
is given as
 \begin{eqnarray}
	f_{\kappa,\beta}=
	\frac{f}{N_k}\sum_{\k}^{}\sum_{n=1}^{occ}
\langle u_{n{\k}}^{(0)}|
(T+v^{}_{ext})^{\tau_{\kappa,\beta}}|u_{n{\k}}^{{(0)}}\rangle \;,
\end{eqnarray}
and taking an additional derivative with respect to electric field yields
\begin{eqnarray}
Z^*_{\kappa,\alpha \beta}=
	\frac{2f}{N_k}\sum_{\k}^{}\sum_{n=1}^{occ}
\langle u_{n{\k}}^{(0)}|
(T+v^{}_{ext})^{\tau_{\kappa,\beta}}|u_{n{\k}}^{\E_{\alpha}}\rangle \;.
\label{eq:zstara}
\end{eqnarray}
This has essentially the same form as Eq.~(43) in
Ref.~\onlinecite{dfpt2}, except that here the zero-order wave functions
are already polarized by the preexisting finite electric field.

Alternatively, Eq.~(\ref{eq:bec}) can be computed as the derivative of
the polarization with respect to the displacement,
\begin{eqnarray}
Z^*_{\kappa,\alpha \beta}=\Omega\frac{\partial
P_{\alpha}}{\partial \tau_{\kappa,\beta}}=\Omega {\hat {\bf e}_{\alpha}}
	\cdot \P^{\tau_{\kappa,\beta}}_{}\;.
\label{eq:zstarb}
\end{eqnarray}
Here $\P^{\tau_{\kappa,\beta}}_{}$ takes a form very similar
to that of Eq.~(\ref{eq:ep1a}), except that the first-order
changes $|u_{n{\k_j}}^{\E_{\alpha}}\rangle$ in the wave functions in
response to an electric field are replaced by the corresponding changes
$|u_{n{\k_j}}^{\tau_{\kappa,\beta}}\rangle$ in response to a sublattice
displacement.  The computation of the
$|u_{n{\k_j}}^{\tau_{\kappa,\beta}}\rangle$ has already been
described in detail in Ref.~\onlinecite{wang06}.

The computation of the first-order derivatives of the wave functions is
typically the most time-consuming step of the linear-response calculation.
Therefore, for a complicated unit cell with many atoms $M$ per cell,
the computation of the three derivatives $|u^{\E_{\alpha}}\rangle$
will be much cheaper than that of the $3M$
derivatives $|u^{\tau_{\kappa,\beta}}\rangle$, and the method of
Eq.~(\ref{eq:zstara}) will therefore be significantly faster than
the method of Eq.~(\ref{eq:zstarb}).  In the special case that the
displacement derivatives $|u^{\tau_{\kappa,\beta}}\rangle$ have
already been computed for some other reason (e.g., for the purpose of
computing the phonon frequencies in finite field), the use of
the latter method may be advantageous.  In any case, a comparison
of the two methods should provide a useful check on the internal
consistency of the theory and its computational implementation.

\section{Test calculations for III-V semiconductors}
\label{sec:test}

In order to check our method, we have performed test calculations
on two prototypical III-V semiconductors, AlAs and GaAs, for which
the electronic contribution to the polarization is typically
comparable to the ionic contribution.\cite{dieguez06}
The calculation is carried out using the planewave-pseudopotential
method based on density-functional theory with local-density
approximation (LDA). We use Troullier-Martins norm-conserving
pseu\-do\-po\-tentials\cite{troullier91} in which the $3d$
states on the Ga and As atoms are treated as core states.
(The omission of the semicore $3d$ states from the valence on
the Ga atom may limit the accuracy of the Ga pseudopotential somewhat.)
A $16\times 16 \times 16$ Monkhorst-Pack
mesh is used for the $k$-point sampling. More computational
details can be found in our preceding paper.\cite{wang06}

The calculation of the dielectric permittivity tensor and the Born
effective charge tensor is carried out in three steps.  First, a
ground-state calculation at finite electric field is performed
using the Berry-phase approach\cite{souza02} implemented in the
{\tt ABINIT} code, and the field-polarized Bloch functions are stored
for the later linear response calculation. Second, the linear response
calculation is carried out to obtain the first-order response of
Bloch functions.  Third, the matrix elements of the dielectric
and Born effective charge tensors are computed using these first-order
responses.

\begin{table}
\caption{Calculated electronic dielectric constants of AlAs and
GaAs at zero field, and changes resulting from an electric field of
$3.08 \times 10^8$ V/m along the [100] direction.  `LR' and `FD'
denote the results of linear-response [Eq.~(\ref{eq:dies})]
and finite-difference calculations, respectively.}
\begin{ruledtabular}
\begin{tabular}{ccrrrr}
& &
\multicolumn{1}{c}{$\epsilon_{\infty}$} &
\multicolumn{1}{c}{$\Delta\epsilon_{\infty,23}$} &
\multicolumn{1}{c}{$\Delta\epsilon_{\infty,11}$} &
\multicolumn{1}{c}{$\Delta\epsilon_{\infty,33}$} \cr
\hline
AlAs & LR &  9.681 & 0.039 & 0.027 & 0.013 \cr
     & FD &  9.681 & 0.040 & 0.027 & 0.013 \cr
GaAs & LR & 13.315 & 0.202 & 0.211 & 0.104 \cr
     & FD & 13.319 & 0.203 & 0.207 & 0.098 \cr
\end{tabular}
\end{ruledtabular}
\label{table:one-new}
\end{table}

The first column of Table~\ref{table:one-new} shows the calculated
electronic dielectric constants of AlAs and GaAs at zero electric
field, and the remaining ones show the nonzero changes in the
dielectric tensor elements after the application of an electric field
$\E^{(0)}$ of $3.08 \times 10^8$ V/m along the [100] direction.
The results obtained with the linear-response approach of
Eq.~(\ref{eq:dies}) are compared with those calculated by finite
differences.  In the latter case, polarizations are computed at
several values of the electric field in steps of $3.08 \times
10^5$ V/m, and the dielectric tensor is calculated using a
finite-difference version of Eq.~(\ref{eq:dies}).  It can be seen
that the agreement between the linear-response and the
finite-difference results is excellent, demonstrating the internal
consistency between the two approaches.

In Table ~\ref{table:two-new} we present similar results for the
cation Born effective charges of the same two materials,
first at zero field and then again under application of a
field of $\E^{(0)}$ of $3.08 \times 10^8$ V/m along the [100]
direction.  The linear-response results were obtained using
Eq.~(\ref{eq:zstara}), but we also computed the corresponding
values using Eq.~(\ref{eq:zstarb}) and found agreement between
the two linear-response approaches with a maximum fractional
error smaller than $10^{-6}$ for all values reported.  For the
finite-difference comparison, the polarizations were computed
at several values of the atomic displacements in steps of
10$^{-3}$ Bohr and the Born charge tensors were calculated using
a finite-difference version of Eq.~(\ref{eq:zstarb}).  It can
again be seen the agreement between the linear-response and the
finite-difference results is excellent.

We emphasize that the values of $\Delta\epsilon_\infty$ and
$\Delta Z^*$ reported in Tables \ref{table:one-new} and
\ref{table:two-new} are purely electronic or ``frozen-ion'' ones
-- that is, the sublattice displacements that would be induced
by a truly static electric field $\E^{(0)}$ are not included.

\begin{table}
\caption{Calculated cation Born effective charges of AlAs and
GaAs at zero field, and changes resulting from an electric field of
$3.08 \times 10^8$ V/m along the [100] direction.  `LR' and `FD'
denote the results of linear-response [Eq.~(\ref{eq:zstara})]
and finite-difference calculations, respectively.}
\begin{ruledtabular}
\begin{tabular}{ccrrrr}
& &
\multicolumn{1}{c}{$Z^*$} &
\multicolumn{1}{c}{$\Delta Z^*_{23}$} &
\multicolumn{1}{c}{$\Delta Z^*_{11}$} &
\multicolumn{1}{c}{$\Delta Z^*_{33}$} \cr
& & & $(\times10^{-3})$ & $(\times10^{-3})$ & $(\times10^{-3})$ \cr
\hline
AlAs & LR &  2.110 & 17.23 & $-$0.06 & $-$0.13 \cr
     & FD &  2.110 & 17.22 & $-$0.05 & $-$0.11 \cr
GaAs & LR &  2.186 & 52.88 & $-$3.42 & $-$3.17 \cr
     & FD &  2.186 & 52.83 & $-$3.36 & $-$3.14 \cr
\end{tabular}
\end{ruledtabular}
\label{table:two-new}
\end{table}

The values of $\epsilon_\infty$ and $Z^*$ reported in Tables
\ref{table:one-new} and \ref{table:two-new} are in good agreement
with other theoretical values in the
literature\cite{baroni87,Gironcoli89,sengstag95}
and with experiment.  The symmetry is such that the applied
electric field along $x$ breaks the degeneracy between the diagonal
elements of the $\epsilon_\infty$ and $Z^*$ tensors so that
$\epsilon_{\infty,11}\neq\epsilon_{\infty,22}=\epsilon_{\infty,33}$
and $Z^*_{11}\neq Z^*_{22}=Z^*_{33}$, and introduces non-zero
off-diagonal elements $\epsilon_{\infty,23}=\epsilon_{\infty,32}$
and $Z^*_{23}=Z^*_{32}$.

Symmetry considerations also imply that
$\epsilon_{\infty,23}$ and $Z^*_{23}$ should appear to first order
in $\E^{(0)}$, while
$\Delta\epsilon_{\infty,11}$,
$\Delta\epsilon_{\infty,33}$,
$\Delta Z^*_{11}$, and
$\Delta Z^*_{33}$
should be quadratic in $\E^{(0)}$.
This is confirmed by our numerical calculations.  Indeed, by
repeating calculations like those shown in Tables
\ref{table:one-new} and \ref{table:two-new} for several values of
$\E^{(0)}$ and fitting to obtain the coefficients of the linear and
quadratic dependence, we can extract information about the
nonlinear dielectric response and the Raman tensor.  The
second- and third-order nonlinear dielectric tensors are defined as
\begin{equation}
    \chi^{(2)}_{1 2 3}
       =\frac{1}{2}\,\frac{\partial^2 P_{2}} {\partial \E_{1}\partial \E_{3}}
       =\frac{1}{2}\,\frac{\partial   \chi_{23}} {\partial \E_{1}}
\label{eq:chi123}
\end{equation}
and
\begin{equation}
    \chi^{(3)}_{1111}
       =\frac{1}{6}\,\frac{\partial^3 P_{1}} {\partial \E_{1}^3}
       =\frac{1}{6}\,\frac{\partial^2 \chi_{11}} {\partial \E_{1}^2} \;,
\label{eq:chi1111}
\end{equation}
while the Raman polarizability tensor is defined by
\begin{equation}
    \alpha_{\rm TO}
       =\,\frac{\partial^2 f_2}{\partial \E_{1}\partial \E_{3}}
       =\,\frac{\partial Z_{23}}{\partial \E_{1}}
\label{eq:raman}
\end{equation}
where $\bf f$ is the force on the cation sublattice induced by the
electric field.  In practice, we calculate $\chi_{23}$, $\chi_{11}$,
and $Z^*_{23}$ for a series of finite electric fields
oriented along the $x$-axis with values of $\E^{(0)}$ ranging from
zero to $5.14 \times 10^8$V/m in increments of one-fifth of the
maximum value.  Fitting these data to a polynomial in $\E^{(0)}$
then gives the values of $\chi^{(2)}_{123}$, $\chi^{(3)}_{1111}$,
and $\alpha_{\rm TO}$.  Note that $\alpha_{\rm TO}$ can
alternatively be expressed as
\begin{equation}
    \alpha_{\rm TO}
       ={\Omega}\,\frac{\partial \chi_{23}}{\partial\tau_{1}}
\label{eq:ramanalt}
\end{equation}
where $\tau_1$ is a cation sublattice displacement and $\chi_{23}$
is computed at zero field.  We have also computed $\alpha_{\rm TO}$
by fitting to a series of calculations of this type, and find values
of $\alpha_{\rm TO}$ that agree with those obtained from Eq.~(\ref{eq:raman})
within $0.3$\%.

\begin{table}
\caption{Values of second-order dielectric susceptibility
and Raman matrix elements in AlAs, as defined by Eqs.~(\ref{eq:chi123}) and
(\ref{eq:raman}) respectively, compared with previous theory and
experiment.}
\begin{ruledtabular}
\begin{tabular}{lcc}
 & $\chi^{(2)}_{123}$ (pm/V) & $|\alpha_{\rm TO}|$ (\AA$^2$) \cr
 \hline
Present work                           & 62 & 8.0 \cr
Theory,\footnotemark[1] Ref.~\onlinecite{souza02}      & 64 &     \cr
Theory,\footnotemark[2] Ref.~\onlinecite{veithen05}    & 70 & 8.5 \cr
Theory,\footnotemark[1] Ref.~\onlinecite{roman06}      & 79 & 9.0 \cr
Theory,\footnotemark[2] Ref.~\onlinecite{deinzer02}    &    & 7.4 \cr
Experiment, Ref~\onlinecite{shoji02} &   78$\pm$20  &  \cr
\end{tabular}
\end{ruledtabular}
\label{table:three-new}
\footnotetext[1]{Using finite-difference approach.}
\footnotetext[2]{Using ($2n+1$)-theorem approach.}
\end{table}

The results for the $\chi^{(2)}_{123}$ and $\alpha_{\rm TO}$
values as computed from Eqs.~(\ref{eq:chi123}) and (\ref{eq:raman})
are presented in Table \ref{table:three-new} for AlAs, together
with some previous theoretical and experimental values for comparison.
In view of the fact that the calculation of higher-order tensor
elements tends to be delicate, the agreement is generally quite good.
In particular, Veithen et al.\cite{veithen05} have shown (see their
Fig.~1) that the results for  $\chi^{(2)}_{123}$ can be quite
sensitive to the method of discretization in $k$-space and the
fineness of the $k$-point mesh.  For GaAs we find
$\chi^{(2)}_{123}= 293$\,pm/V and $\alpha_{\rm TO}=-24.1$\,\AA$^2$ 
(which is close to the value in Ref.~\onlinecite{deinzer02}),
but these numbers are of questionable accuracy because of
our use of a Ga pseudopotential that does not include the $3d$ semicore
orbitals in the valence.
We obtain $\chi_{1111}^{(3)}$ values of 3.90 and 33.8$\times10^{-11}$\,esu
for AlAs and GaAs, respectively.
We are not aware of previous theoretical values of $\chi_{1111}^{(3)}$
with which to compare; this quantity is beyond the reach of the
``$2n+1$'' theorem using first-order wave function responses only,
and so is difficult to compute by pure DFPT methods.  Experimental
values ranging from 3.9 to 18$\times10^{-11}$\,esu for GaAs\cite{ching93}
can be found in the literature.

The discrepancies noted above between theory and theory, and between
theory and experiment, may have many possible causes.  In addition
to some of the computational and convergence issues mentioned above,
the adequacy of the LDA approximation itself is also a
serious question. Because the LDA tends to underestimate gaps, some authors
have included a so-called ``scissors correction'' in order to
widen the gap artificially; this tends to decrease the magnitude
of response tensors.\cite{aulbur96}
On the experimental side, the difficulty in obtaining reproducible
results is surely also an issue.  Nevertheless, we emphasize
that the {\it relative} accuracy of the values reported in Tables I and
II, which were done under the {\it same} computational conditions (same
pseudopotentials, $k$-point meshes, etc.), demonstrates the correctness
of our new finite-field linear-response formulation and the internal
consistency of the computational framework that we employ.

\section{Summary}
\label{sec:sum}

We have developed a linear-response method for
computing dielectric constants and Born effective charges in the
presence of a {\it finite} electric field.  We have demonstrated
the reliability of our approach by implementing it in the context
of the {\tt ABINIT} code package\cite{abinit} and performing test
calculations on two III-V semiconductors, AlAs and GaAs.  We have
confirmed that the results calculated using the new linear-response
approach are consistent with those obtained from finite-difference
calculations carried out within the same framework.  In general,
our results are also in good agreement with other theoretical
calculations and with experiment.

A major advantage of the present approach is that, unlike the conventional
long-wave linear-response method,\cite{dfpt1} it can be applied to obtain
response tensors in finite electric field.  While it is possible to
obtain similar information from a set of finite-difference
calculations carried out for some chosen set of applied electric
fields, the linear-response approach is more direct, and it avoids
the troublesome truncation errors that may arise in a
finite-difference approach.  In the future, it may be of interest
to extend the finite-field DFPT treatment not just to phonon
perturbations (presented in Ref.~\onlinecite{wang06}) and
electric-field perturbations (presented here), but also to other
perturbations such as those associated with strain or chemical
composition.  Taken together, these developments should allow for
much greater flexibility in the calculation of materials properties
of insulators under electrical bias and facilitate the study of
higher-order nonlinear dielectric properties.

\acknowledgments

This work was supported by NSF Grant DMR-0549198 and ONR Grant
N00014-05-1-0054.



\begin{thebibliography}{10}

\bibitem{lst} G. Grosso and G.P. Parravicini,
{\it Solid State Physics} (Academic Press 2003) {\bf P} 331-333.

\bibitem{sai02} N. Sai, K.M. Rabe, D. Vanderbilt, Phys. Rev. B {\bf
66}, 104108 (2002).

\bibitem{fu03} H. Fu and L. Bellaiche, Phys. Rev. Lett. {\bf 91},
57601 (2003).

\bibitem{dieguez06} O. Di\'eguez and D. Vanderbilt,
Phys. Rev. Lett. {\bf 96}, 056401 (2006).

\bibitem{antons05} A. Antons, J.B. Neaton, K.M. Rabe, and D. Vanderbilt,
Phys. Rev. B {\bf 71}, 024102 (2005).

\bibitem{wu06} X. Wu, O. Di\'eguez, K.M. Rabe, and D. Vanderbilt,
Phys. Rev. Lett. {\bf 97}, 107602 (2006).

\bibitem{wang06} X. Wang and D. Vanderbilt, Phys. Rev. B {\bf 74},
054304 (2006).

\bibitem{dfpt1} X. Gonze, Phys. Rev. B {\bf 55}, 10337 (1997).

\bibitem{dfpt2} X. Gonze and C. Lee, Phys. Rev. B {\bf 55}, 10355 (1997).

\bibitem{nunes01} R.W. Nunes and X. Gonze, Phys. Rev. B {\bf 63},
155107 (2001).

\bibitem{souza02} I. Souza, J. \'I\~niguez, and D. Vanderbilt,
Phys. Rev. Lett. {\bf 89}, 117602 (2002).

\bibitem{umari02} P. Umari and A. Pasquarello, Phys. Rev.
Lett. {\bf 89}, 157602 (2002).

\bibitem{souza04} I. Souza, J. \'I\~niguez, and D. Vanderbilt,
Phys. Rev. B {\bf 69}, 085106 (2004).

\bibitem{abinit}
The ABINIT code is a common project of the Universit\'e
Catholique de Louvain, Corning Incorporated, and other
contributors (www.abinit.org). See: X. Gonze and others,
Comp. Mat. Science {\bf 25}, 478 (2002).

\bibitem{wu05} X. Wu, D. Vanderbilt, and D.R. Hamann,
Phys. Rev. B {\bf 72}, 035105 (2005).

\bibitem{troullier91} N. Troullier and J.L. Martins,
Phys. Rev.  B {\bf 43}, 1993 (1991).

\bibitem{baroni87} S. Baroni, P. Giannozzi and A. Testa, Phys. Rev. Lett.
{\bf 58}, 1861 (1987).

\bibitem{Gironcoli89}S. Gironcoli, S. Baroni and R. Resta,
Phys. Rev. Lett. {\bf 62}, 2853  (1989).

\bibitem{sengstag95}T. Sengstag, N. Binggeli, and A. Baldereschi,
Phys. Rev. B {\bf 52}, R8613 (1995).


\bibitem{veithen05} M. Veithen, X. Gonze, and Ph. Ghosez,
Phys. Rev. B {\bf 71}, 125107 (2005).

\bibitem{roman06} E. Roman, J.R. Yates, M. Veithen, D. Vanderbilt
and I. Souza, Phys. Rev. B {\bf 74}, 245204 (2006).

\bibitem{deinzer02} G. Deinzer and D. Strauch, Phys. Rev. B {\bf 66}, 100301 (2002).

\bibitem{shoji02} I. Shoji, T. Kondo and R. Ito, Optical and
Quantum Electronics {\bf 34} 797, (2002).

\bibitem{ching93} W.Y. Ching and Ming-Zhu Huang, Phys. Rev. B {\bf 47},
9479 (1993).




\bibitem{aulbur96} W.G. Aulbur, L. J\"onsson, and J.W. Wilkins, Phys.
Rev. B {\bf 54}, 8540 (1996).

\end{thebibliography}
\end{document}